\documentclass[a4paper,fleqn,usenatbib]{mnras}

\usepackage[T1]{fontenc}
\usepackage{ae,aecompl}
\usepackage{graphicx}	
\usepackage{amsmath}	
\usepackage{amssymb}	
\usepackage{nicefrac} 
\usepackage{booktabs}
\usepackage[none]{hyphenat}
\usepackage{tabularx}   

\title[Identifying the brightest Galactic globular clusters]{Identifying the brightest Galactic globular clusters for future observations by H.E.S.S.\ and CTA}

\author[]{
Hambeleleni Ndiyavala$^{1,2}$\thanks{E-mail: 26403366@nwu.ac.za},
Petrus Paulus Kr\"{u}ger$^{1}$,\newauthor
and Christo Venter$^{1}$\newauthor \\
\\
$^{1}$Centre for Space Research, North-West University, Potchefstroom Campus, Private Bag X6001, Potchefstroom, 2520, South Africa\\
$^{2}$University of Namibia, Khomasdal Campus, Private Bag 13301, Windhoek, Namibia
}

\date{Accepted XXX. Received YYY; in original form ZZZ}

\pubyear{2017}

\begin{document}
\label{firstpage}
\pagerange{\pageref{firstpage}--\pageref{lastpage}}
\maketitle

\begin{abstract}
We present results from an emission code that assumes millisecond pulsars (MSPs) to be sources of relativistic particles in globular clusters (GCs) and models the resulting spectral energy distribution (SED) of Galactic GCs due to these particle's interaction with the cluster magnetic and soft-photon fields. We solve a transport equation for leptons and calculate inverse Compton (IC) and synchrotron radiation (SR) to make predictions for the flux expected from Galactic GCs. We perform a parameter study and also constrain model parameters for three GCs using $\gamma$-ray and X-ray data. We next study the detectability of 16 Galactic GCs for the High Energy Stereoscopic System (H.E.S.S.) and the Cherenkov Telescope Array (CTA), ranking them according to their predicted TeV flux. The spectrum of each cluster and therefore the detectability ranking is very sensitive to the choice of parameters. We expect H.E.S.S.\ to detect two more GCs (in addition to Terzan~5), i.e., 47 Tucanae and NGC~6388, if these clusters are observed for 100 hours. The five most promising GCs for CTA are NGC~6388, 47~Tucanae, Terzan~5, Djorg~2, and Terzan~10. We lastly expect CTA to detect more than half of the known Galactic GCs population, depending on observation time and model parameters.
\end{abstract}

\begin{keywords}
globular clusters: general - pulsars: general - radiation mechanisms: non-thermal
\end{keywords}

\section{Introduction}\label{Introduction}
Globular clusters (GCs) are among the most ancient bound stellar systems in the Universe, consisting of $10^{4} - 10^{6}$ stars (e.g., \citealt{Lang1992}). They are normally associated with a host galaxy and most galaxies, including the Milky Way, are penetrated and surrounded by a system of GCs. There are nearly 160 known Galactic GCs \citep{Harris2010}, and they are spherically distributed about the Galactic Centre lying at an average distance of $\sim 12$~kpc.

The \emph{Fermi} Large Area Telescope (LAT) has detected about a dozen GCs \citep{Nolan2012}. The \emph{Astro Rivelatore Gamma a Immagini Leggero (AGILE)}, which is an X-ray and $\gamma$-ray instrument, has however not detected any GC to date due to its lower sensitivity. The ground-based Cherenkov telescope, the High Energy Stereoscopic System (H.E.S.S.), which is operated in a pointing mode (limiting the fraction of the sky it can annually observe) has only detected a single cluster within our Galaxy, i.e., Terzan~5 at very-high energies \citep[VHEs; $> 100$~GeV][]{Abramowski2011}. Other Cherenkov telescopes could only produce upper limits (e.g., \citealt{Anderhub2009}). The future Cherenkov Telescope Array (CTA) will be about 10 times more sensitive than H.E.S.S.\ \citep{Wagner2010} and is expected to detect TeV emission from a few more GCs. In addition, diffuse radio (e.g., \citealt{Clapson2011}) and diffuse X-ray emission (e.g., \citealt{Eger2010,Eger2012,Wu2014}) have also been detected from some GCs. 

On the other hand, radio, X-ray, and high-energy (HE) $\gamma$-ray pulsars have been detected in some GCs. For example, the bright X-ray GC pulsar B1821$-$24 \citep{Hui2009} was found in M28 via radio observations \citep{Lyne1987}, making this the first ever pulsar later detected in a GC. Significant $\gamma$-ray pulsations have since been detected from it \citep{Johnson2013}. \citet{Bogdanov2011} detected X-ray pulsations as well as a non-thermal spectrum from PSR B1824$-$21, which is probably due to magnetospheric emission, an unresolved pulsar wind nebula, or small-angle scattering of the pulsed X-rays by interstellar dust grains. Furthermore, \citet{Forestell2014} detected X-rays from four faint radio milisecond pulsars (MSPs) in NGC 6752, consistent with thermal emission from the neutron star surfaces using \textit{Chandra} data. Also the young, energetic MSP PSR~J1823$-$3021A, which has a period of 5.44 ms and an unusually high inferred surface magnetic field of $4\times10^{9}$ G was discovered by \citet{Biggs1994} in a survey of GCs and then detected by \emph{Fermi} in NGC~6624 \citep{Freire2011}. Another example of point sources in GCs is afforded by the 25 radio MSPs in 47 Tucanae (see \citealt{Pan2016,Ridolfi2016,Freire2017} for recent updates), 19 of which have been identified at X-rays \citep{Heinke2005,Bogdanov2006}. See \citet{Ransom2008, Freire2013, Hessel2015, Hui2009} for an overview of radio and X-ray pulsars in GCs.

The pulsars are not detectable at optical wavelengths among the stellar emission in the cluster, but some of their companions are (for some recent examples see, e.g., \citealt{Pallanca2010, Pallanca2013, Mucciarelli2013, Cadelano2015, Rivera2015, Pallanca2016}), together with many other types of exotic binaries: cataclysmic variables, active binaries, and blue stragglers (e.g., \citealt{Geffert1997,Heinke2003, Heinke2005}). GCs thus form an important class of Galactic emitters of broadband radiation, both point-like and diffuse in morphology, and are prime targets for deeper future observations by more sensitive telescopes.

Several models exist that predict the multi-wavelength spectrum radiated by GCs. \citet{Bednarek2007} considered a scenario where leptons are accelerated by MSPs at relativistic shocks that are created when their winds collide with each other inside the cores of these clusters. These leptons upscatter ambient photons via inverse Compton (IC) scattering, which may lead to unpulsed GeV to TeV spectral emission components. \citet{Harding2005} furthermore modelled the cumulative pulsed GeV flux via curvature radiation (CR) from MSP magnetospheres by assuming a pair-starved polar cap electric field \citep[e.g.,][]{Venter2005}. \citet{Venter2008a} modelled the cumulative pulsed CR from 100 such pulsars by randomising over MSP geometry as well as period and period time derivative. \citet{Venter2010} refined this approach and could predict the GeV spectrum of 47 Tucanae within a factor of two in both energy and flux level, prior to its detection by \emph{Fermi} LAT. \citet{Cheng2010} considered an alternative scenario to produce GeV emission and calculated unpulsed IC radiation from electrons and positrons upscattering the cosmic microwave background (CMB), stellar photons, and the Galactic background. This is in contrast to the usual assumption that the GeV emission measured by \emph{Fermi} is due to pulsed CR. \citet{Bednarek2016} further developed a scenario in which they considered both the diffusion process of leptons in a GC and also their advection by the wind produced by the mixture of winds\footnote{This type of intracluster medium (ICM) has been detected in 47 Tucanae \citep{Freire2001}} from the population of MSPs and red giant stars within the GC. They also considered the spatial distribution of MSPs within a GC and the effects related to the non-central location of an energetic, dominating MSP. Finally, there is a hadronic model that attempts to explain the observed TeV emission from Terzan~5 \citep{Domainko2011}. For a review, see \citet{Bednarek2011, Tam2016}.

In this paper, we use a multi-zone, steady-state, spherically symmetric model \citep{Kopp2013}, based on the work of \citet{Venter2009a}, that calculates the lepton transport (including diffusion and radiation losses) and predicts the spectral energy distribution (SED) from GCs for a very broad energy range by considering synchrotron radiation (SR) as well as IC emission. This model assumes the so-called `MSP scenario' where MSPs are thought to be responsible for the relativistic particles that emit both pulsed CR and unpulsed (SR and IC) emission. This paper therefore represents an application of the \citet{Kopp2013} model and reports on the detectability of Galactic GCs with H.E.S.S.\ and CTA by using this model to systematically estimate the flux of all Galactic GCs.
The rest of the paper is structured as follows. In Section~\ref{model}, we discuss the transport equation solved in \citet{Kopp2013} to obtain the steady-state electron spectrum, including the assumed particle injection spectrum, diffusion coefficients, radiation loss terms, and soft-photon target fields. In Section~\ref{Parameters}, we perform a parameter study to investigate the model's behaviour and to study the degeneracy between free parameters. Section~\ref{Spectral} includes a discussion on the parameters of three GCs that we have constrained using multi-wavelength data and also a list of the five most promising GCs for CTA based on their predicted VHE flux. We offer our outlook in Section~\ref{Conclusion}.

\section{The model}
\label{model}
In this paper, we use a model by \citet{Kopp2013} that calculates the particle transport and observed spectrum for GCs. Below, we summarise the major aspects of this model.

A Fokker-Planck-type equation \citep{Parker1965} prescribes the transport of relativistic electrons and positrons in GCs. Neglecting spatial convection, we have \citep{Kopp2013}:
\begin{equation}
 \frac{\partial n_{\rm e}}{\partial t} = \boldsymbol{\nabla} . (\boldsymbol{\kappa} . \boldsymbol{\nabla} n_{\rm e}) - 
\frac{\partial}{\partial E_{\rm e}}(\dot E_{\rm e}n_{\rm e}) + Q_{\rm tot},
\label{eq:1}
\end{equation}
where $n_{\rm e}$ is the electron density per energy and volume and is a function of central GC radius $\mathbf{r}_{\rm s}$, $E_{\rm e}$ the electron energy, $\boldsymbol{\kappa}$ is the diffusion tensor, $\dot E_{\rm e}$ the radiation losses, and $Q$ the electron source term.

Since H.E.S.S.\ detected a power-law spectrum for the VHE source associated with Terzan~5, we assume that the particle injection spectrum is also a power law (see Eq.~[\ref{eq:5}]) between energies $E_{\rm e, \rm min}~{\rm and}~E_{\rm e, \rm max}$. Assuming the source term $Q$ is located at $r_{\rm s}~=~r_{\rm inj}$ and follows a power-law distribution, we have
\begin{equation}
 Q_{\rm tot} = Q_{0}\frac{\delta(r_{\rm s}-{r}_{\rm inj})}{E_{\rm e}^{\Gamma}} = \sum\limits_{i=1}^{N_{\rm MSP}}Q_{i},
 \label{eq:5}
\end{equation}
where $\Gamma$ is the spectral index, $Q_0$ is the normalisation constant, and $N_{\rm MSP}$ the number of MSPs in the GC. The normalisation of the injection spectrum requires
\begin{align}\nonumber
L_{\rm e}\equiv\sum\limits_{i=1}^{N_{\rm MSP}}\int_{E_{\rm e,\rm min}}^{E_{\rm e,\rm max}}E_{\rm e}Q_{i}\,dE_{\rm e}&=N_{\rm MSP}\eta\langle \dot E_{\rm rot}\rangle \\
&=\int_{E_{\rm e,\rm min}}^{E_{\rm e,\rm max}}E_{\rm e}Q_{\rm tot}\,dE_{\rm e},
\label{eq:10}
\end{align}
with $\eta$ the fraction of the average MSP spin-down power $\dot E_{\rm rot}$ converted into particle power \citep{Bednarek2007}.
Using the Gauss theorem and assuming $\dot E_{\rm e} = 0$ for the innermost zone, the source term $Q_{\rm tot}$ may be replaced by a boundary condition:
\begin{equation}
\left.\frac{\partial n_{\rm e}}{\partial r_{\rm s}} \right\vert_{r_{\rm s, min}} = -\frac{Q_{0}}{4\pi r^{2}_{\rm s, min}\kappa (E_{\rm e})E_{\rm e}^{\Gamma}},
\end{equation}
with $r_{\rm min}$ bounding the spherical region containing all particle 
sources.
While we duly note the non-asymmetric source distribution of MSPs in GCs (e.g., in some GCs like 47 Tucanae, MSPs are located within a region with a radius of 1.7 pc, which is larger than the core radius) for simplicity this model implements spherical symmetry. This forces us to assume that the pulsars are located in the centre of the GC (within a radius $r_{\rm min} = 0.01$~pc) as a first approximation. The work of \citet{Bednarek2016} studies the effect of MSPs occurring at different positions in a GC on the predicted GeV and TeV flux. 

We assumed two different diffusion coefficients. First, for Bohm diffusion we have
\begin{equation}
\kappa(E_{\rm e}) = \kappa_{B}\frac{E_{\rm e}}{B}, 
\label{eq:7}
\end{equation}
where $\kappa_{B} = c/3e$, with $c$ being the speed of light, $e$ the elementary charge, and $B$ is the cluster magnetic field. We assumed that $B$ is a constant. Second, we also consider a diffusion coefficient inspired by Galactic cosmic-ray propagation studies,
\begin{equation}
\kappa(E_{\rm e}) = \kappa_{B}\left(\frac{E_{\rm e}}{E_{0}}\right)^{\alpha},
\label{eq:8}
\end{equation}
with $E_{0}=1~\rm TeV~\rm and~\alpha=0.6$, (e.g., \citealt{Moskalenko1998}).

For IC scattering and SR, one needs to specify the energy losses ($\dot E_{\rm e}$). We follow \citet{Kopp2013} by writing the IC losses in the general case (including both the Thomson and Klein-Nishina limits; \citealt{Blumenthal1970}) as
\begin{equation}
\dot E_{\rm IC}(E_{\rm e}, r_{\rm s}, T_{j}) = -\sum_{j=0}^{k-1}\int n_{\varepsilon, j}(r_{\rm s}, \varepsilon, T_{j})\frac{E_{\gamma}}{E_{0}}\zeta(E_{\rm e}, E_{\gamma},\varepsilon)\,d\varepsilon,
\label{eq:11}
\end{equation}
where $k$ is the total number of (blackbody) soft-photon components, $n_{\varepsilon, j}$ the photon density of the $j^{\rm th}$ blackbody emitter, $\varepsilon$ the original photon energy, $T_{j}$ is the blackbody temperature of component $j$, $E_{\gamma}$ represents the final energy of the upscattered photons, and $E_{0}$ the electron rest energy. The collision rate $\zeta$ is defined in \citep{Jones1968}. 
One needs to specify the soft-photon densities in order to calculate $\dot E_{\rm IC}.$
For a blackbody, \citet{Kopp2013} used a photon density (see also \citealt{Zhang2008}):
\begin{equation}
 n_{\varepsilon, j}(r_{\rm s}, \varepsilon, T_{j}) = \frac{15u_{{\rm rad}, 
j}(r_{\rm s}, T_{j})}{(\pi k_{B}T_{j})^4}\frac{\varepsilon^{2}}{e^{\frac{\varepsilon}{k_{\rm B}T_{j}}}-1},
\label{eq:15}
\end{equation}
where $u_{{\rm rad},j}$ is the energy density of the soft photons and $k_{\rm B}$ is Boltzmann's constant. 
\citet{Kopp2013} used CMB, stellar photons, and the Galactic background radiation field at the position of the GC. For the stellar-photon component they used a line-of-sight integration (see \citealt{Bednarek2007,Prinsloo2013})
\begin{gather}\nonumber
n_{\varepsilon, 1}(r_{\rm s}, \varepsilon, T_{1}) = \frac{8\pi}{h^{3}c^{3}}\frac{\varepsilon^{2}}{e^{\frac{\varepsilon}{k_{\rm 
B}T_{1}}}-1}\left(\frac{1}{2}\frac{N_{\rm tot}R_{\star}^{2}}{R_{\rm c}^{2}\widetilde{R}}\right) 
\int_{r^{\prime}=0}^{r^{\prime}=R_{t}}\hat\rho(r^{\prime})\\
\times\frac{r^{\prime}}{r_{\rm s}}~{\rm ln}~\left(\frac{\mid r^{\prime} + r_{\rm s}\mid}{\mid r^{\prime} - 
r_{\rm s}\mid}\right)\,dr^{\prime},
\label{eq:16} 
\end{gather}
where~$h$ is Planck's constant, $N_{\rm tot}$~represents the total number of cluster stars, which can be written as $N_{\rm tot}={M_{\rm tot}}/{\-m}$, with~$M_{\rm tot}$ the total mass of the cluster and~$\-m$ the average stellar mass. Here, $R_{\star}$~is the 
average stellar radius, $R_{\rm c}$~indicates the core radius of the cluster, and~ 
$\widetilde{R} = 2R_{\rm h}-2R_{\rm c}/3 - {R_{\rm h}^{2}}/{R_{\rm t}}$,
with $R_{\rm h}$ the half-mass radius and $R_{\rm t}$~the tidal radius of the cluster. \citet{Kopp2013} used solar values for $R_{\star}~{\rm and}~ m$, assuming that all stars in the simulation have a solar radius and temperature $T = 4~500~K$. Also, $\hat\rho(r^{\prime})$ is the normalised density profile of the cluster stars as used by \citet{Bednarek2007} based on the Michie-King model. In this study we used only the CMB and the stellar photon field.

In the SR case, the loss rate (averaged over all pitch angles) is given by \citep{Blumenthal1970}
\begin{equation}
 \dot E_{\rm SR}(E_{\rm e}, r_{\rm s}) = -\frac{g_{\rm SR}}{8\pi}E_{\rm e}^{2}B^{2}(r_{\rm s}),
 \label{eq:14}
\end{equation}
with $g_{\rm SR} = 4\sigma_{T}c/3E_{0}^{2} = 32\pi(e/E_{0})^{4}c/9$, and where~$\sigma_{T}$ denotes the Thomson cross section. We follow and use the expressions in \citet{Kopp2013} (valid for isotropic electron and photon distributions) to calculate SR and IC scattering emissivities. We follow \citet{Kopp2013} to calculate a 2D projected sky map of radiation from the 3D emitting GC.

\section{Parameter study}
\label{Parameters}
In this Section, we perform a parameter study to investigate the GC model's behaviour upon varying six free parameters (see~Table~\ref{parameter_table}). We use parameters that resemble those of Terzan~5. We also study the degeneracy between the parameters with respect to the predicted flux. 
\subsection{Timescales and Reference Model}
We calculate different timescales to study the dominant transport processes as a function of GC radius $r_{\rm s}$ and electron energy $E_{\rm e}$.
The diffusion timescale is given by
\begin{equation}\label{diff}
t_{\rm diff} = \frac{r_{\rm s}^{2}}{2\kappa}.
\end{equation}

The total radiation timescale is \citep[e.g.,][]{Venter2008b}
\begin{equation}
t_{\rm rad} = \frac{E_{\rm e}}{\dot E_{\rm SR} + \dot E_{\rm IC}}.
\end{equation}
The effective timescale is then given by \citep{Zhang2008}:
\begin{equation}
t_{\rm eff}^{-1} \approx t_{\rm diff}^{-1} + t_{\rm rad}^{-1}.
\end{equation}

In this Section we use the parameters listed in Table~\ref{parameter_table} to calculate a reference model for the timescales, particle spectrum, and SED graphs (using the structural parameters of Terzan~5; see Section~\ref{sec:Characteristics_GCs}). We fix all parameters for the reference model but then just change one parameter at a time as indicated in the subsequent sections. We use $\eta = 10\%$ for all graphs, i.e., the fraction of spin-down power that is converted to particle power.

\begin{table*}
	\caption{In this Table we list the reference model parameter values that resemble those of Terzan~5}.
	\label{parameter_table}     
	\centering   
	\begin{tabular}{|c|c|}        
		\hline
		Parameters & Parameter values \\ 
		\hline
		Diffusion coefficient ($\kappa$)    		& Bohm diffusion \\ 
		Magnetic field ($B$)                		& $5\,\mu\rm G$ \\
		Injection spectral index ($\Gamma$)             & 2.0 \\
		Number of stars	($N_{\rm tot}$)	                & $4.6\times10^{5}$ \\
		Injection spectral normalisation ($Q_{0}$) 	& $6.33\times10^{33}{\rm erg^{-1}\,{s^{-1}}}$ \\
		Distance ($d$) 			 		& $5.9\,{\rm kpc}$ \\
		Core radius ($R_{\rm c}$)             & $0.21\,{\rm pc}$ \\
		Half-mass radius ($R_{\rm hm}$)       & $0.94\,{\rm pc}$ \\
		Tidal radius ($R_{\rm t}$)            & $17.4\,{\rm pc}$ \\
\hline                                             
\end{tabular}
\end{table*}

\begin{figure}
   \centering
   \includegraphics[height=5.5cm, width=8cm]{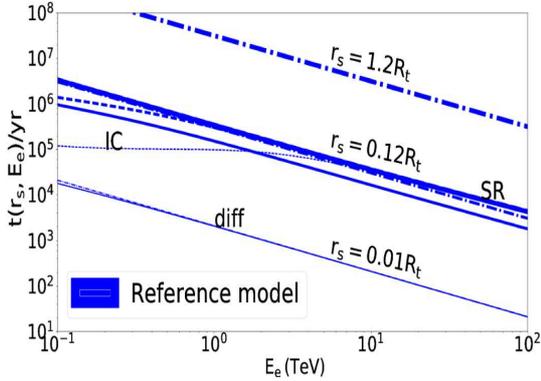}
 \caption{Timescale graph (diffusion indicated by dash-dotted lines, radiation losses by dashed lines, and the effective scale by solid lines) for the reference model as a function of $r_{\rm s}$ and $E_{\rm e}$. Thicker lines indicate larger radii. The labels `IC', `SR', and `diff' indicate where IC, SR, and diffusion dominates, respectively.}
   \label{fig:tls_ref_energy}
 \end{figure}

In Figure~\ref{fig:tls_ref_energy} we plot the timescales for the reference model, indicating diffusion (dash-dotted lines), radiation losses (dashed lines), and the effective timescale (solid lines) as a function of $E_{\rm e}$ for different values of $r_{\rm s}$ (with larger values of $r_{\rm s}$ indicated by thicker lines). The IC cross section drops as one goes from the Thomson regime at low energies to the Klein-Nishina regime at high energies and therefore SR dominates over IC at the highest energies.
We note that both $t_{\rm diff}$ and $t_{\rm SR} \equiv E_{\rm e}/\dot {E}_{\rm SR}$ scale as $E^{-1}$ as seen in Figure~\ref{fig:tls_ref_energy}. Close to the core, diffusion dominates (i.e., particles will escape from a particular zone before radiating). At larger radii the situation is reversed and SR losses dominate over diffusion (since the diffusion timescale scales as $r_{\rm s}^{2}$). At intermediate radii, one can see the change in regime: for $r_{\rm s}= 0.12R_{\rm t}$, with $R_{\rm t}$ the tidal radius, the SR timescale is only slightly lower than the diffusion timescale at the highest particle energies, while SR dominates diffusion at $r_{\rm s}=1.2R_{\rm t}$.

\begin{figure}
  \centering
  \includegraphics[height=5.5cm, width=8cm]{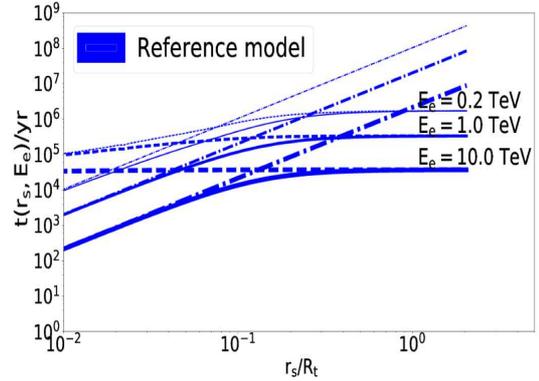}
  \caption{Timescale graph (diffusion indicated by dash-dotted lines, radiation losses by dashed lines, and the effective scale by solid lines) for the reference model as a function of $E_{\rm e}$ and $r_{\rm s}$. Thicker lines indicate larger energies.}
  \label{fig:tls_ref_radius}
\end{figure}

As an alternative view, in Figure~\ref{fig:tls_ref_radius} we plot the timescales as a function of radius for diffusion (dash-dotted lines), radiation losses (dashed lines), and the effective timescale (solid lines) for different energies indicated by different line thickness. For diffusion, the graph of $t_{\rm diff}$ versus $r_{\rm s}$ has a slope of 2 (see Eq.~[\ref{diff}]). Also, $t_{\rm diff}$ is higher for lower energies, since such particles diffuse slower (this is evident at smaller $r_{\rm s}$, where diffusion dominates). At higher particle energies, SR dominates except at the very core. Since we assume that the cluster $B$-field is not a function of $r_{\rm s}$, the graph of $t_{\rm SR}$ versus $r_{\rm s}$ will be flat for constant $E_{\rm e}$. However, $t_{\rm SR}$ is larger for lower particle energies since $t_{\rm SR}~\propto~E_{\rm e}^{-1}$, while IC implies a lower $t_{\rm rad}$ at low ${E_{\rm e}}$ and ${r_{\rm s}}$. From the plot, it is clear that the effective timescale `takes the minimum' between $t_{\rm diff}$ and $t_{\rm SR}$ (or $t_{\rm rad}$), always being the shortest timescale. This is what determines the effect of particle transport on $n_{\rm e}$. One can again see that diffusion dominates radiation processes at low radii. The inverse is true at larger radii. 

\begin{figure}
   \centering
   \includegraphics[height=5.5cm, width=8cm]{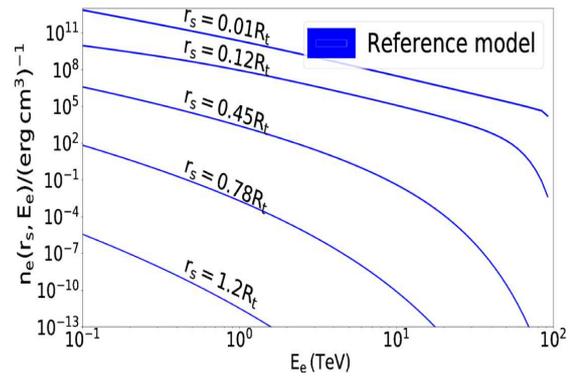}
  \caption[The steady-state particle spectrum as a function of energy for changing $N_{\rm tot}$]{The steady-state particle spectrum as a function of energy $E_{\rm e}$~at different radii $r_{\rm s}$.}
   \label{fig:mel_ref_energy}
 \end{figure}

In Figure~\ref{fig:mel_ref_energy} we plot the steady-state particle spectrum as a function of energy $E_{\rm e}$ at different $r_{\rm s}$ for the reference model. At a fixed radius, $n_{\rm e}$ is higher at low energies and becomes low at higher energies due to the assumed injection spectrum. At small radius, the particle spectrum follows the injection spectrum ($\Gamma=2$). The particle density drops with distance (due to the increased volume).
A spectral cutoff is introduced at higher energies due to SR losses. The cutoff becomes increasingly lower at larger radii.

\begin{figure}
   \centering
   \includegraphics[height=5.5cm, width=8cm]{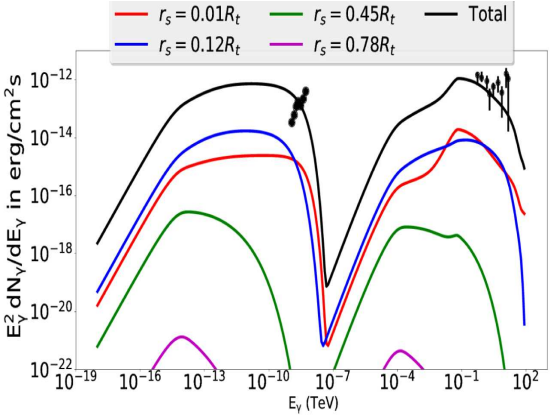}
   \caption{We plot the SED for the reference model (black line) for Terzan~5, as well as \textit{Chandra} \citep{Eger2010} and H.E.S.S.\ \citep{Abramowski2011} data. The relative contributions for a number of representative radii (zones) are also shown using different colours.}
   \label{fig:dsp_ref}
 \end{figure}

Figure~\ref{fig:dsp_ref} shows the SED predicted by the reference model (black curve). There are two components: SR and IC. For the IC component one can observe two sub-components associated with the two soft-photon target densities. We also plot \textit{Chandra} \citep{Eger2010} and H.E.S.S.\ \citep{Abramowski2011} data for Terzan~5. We plot the relative contributions from a number of zones at representative radii $r_{s}$ indicated by different colours. Initially the contribution grows with radius (due to the increased volume of the zones) but farther out drops significantly due to a decline in soft-photon and particle densities. The high-energy part of the IC component dominates the SR component at very small distances where the stellar soft-photon background is larger, while the SR one starts to slightly dominate the IC component at larger distances. 

\subsection{Changing the number of stars ($N_{\rm tot}$)}
\begin{figure}
   \centering
   \includegraphics[height=5.5cm, width=8cm]{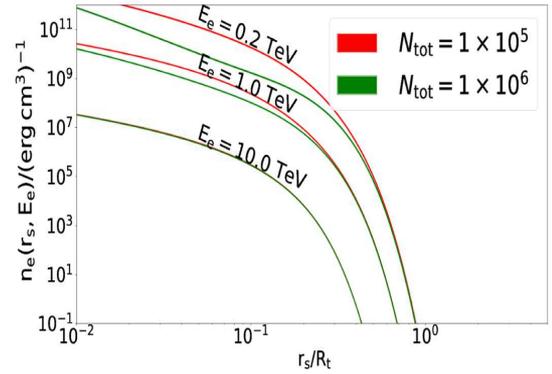}
   \caption{The steady-state particle spectrum as a function of radius $r_{\rm s}$ at different particle energies $E_{\rm e}$.}
   \label{fig:mel_stars_radius}
 \end{figure}

The number density of soft photons $n_{\varepsilon}$ scales linearly with $N_{\rm tot}$ (see Eq.~[\ref{eq:16}]). Thus the IC loss rate also scales linearly with $N_{\rm tot}$ (see Eq.~[\ref{eq:11}]). For a smaller number of stars the photon number density is lower and hence the IC loss rate is lower. It therefore takes a longer time for the particles to lose energy in this case. This effect is evident at lower particle energies and smaller radii (as measured from the cluster centre) where IC dominates over SR due to the high value of $n_{\varepsilon}$ there, leading to a relatively larger $n_{\rm e}$ (Figure~\ref{fig:mel_stars_radius}). At larger radii, the photon number density rapidly declines (leading to smaller $\dot{E}_{\rm IC}$~and longer $t_{\rm IC}$); thus SR (which is not a function of $N_{\rm tot}$) dominates over IC (assuming a constant $B$-field, and therefore the graphs of $n_{\rm e}$ versus $r_{\rm s}$ coincide).
Furthermore, the overall level of $n_{\rm e}$ decreases with radius since it represents a particle density, and the volume scales as $r_{\rm s}^{3}$. 

\begin{figure}
   \centering
   \includegraphics[height=5.5cm, width=8cm]{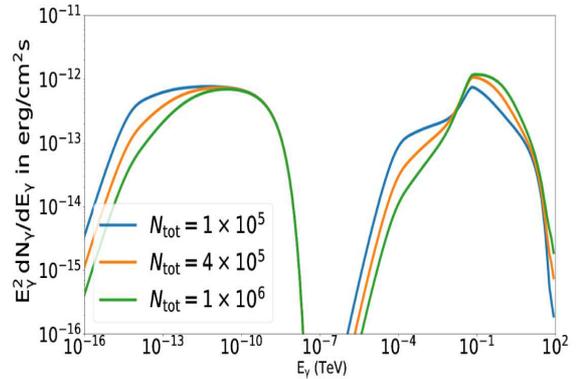}
  \caption{SED plot indicating the effect of changing the number of stars in the cluster.}
  \label{N}
 \end{figure}

Figure~\ref{N} shows that when one increases $N_{\rm tot}$, there are more optical photons, which boosts the IC emission and loss rate. Thus there will fewer low-energy particles and SR and IC on the CMB will be suppressed.

\subsection{Changing the $B$-field}
\begin{figure}
    \centering
    \includegraphics[height=5.5cm, width=8cm]{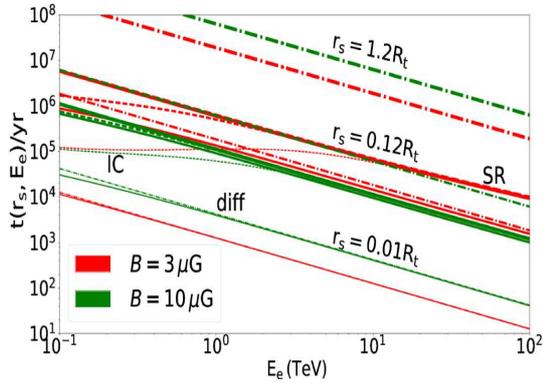}
  \caption{Same as Figure~\ref{fig:tls_ref_energy} but for two different $B$-fields.}
    \label{fig:tlsB_E}
  \end{figure}

We do not know the magnetisation state of plasma in GCs observationally. However we can use reasonable values of $B\sim1-10\,\mu$G. This estimate is due to \citet{Bednarek2007} using typical values for the pulsar wind shock radius and magnetisation parameter $\sigma$ (ratio of the Poynting to particle energy flux). These values also yield reasonable SR and IC spectra.
One expects two main effects when changing the $B$-field: A larger $B$-field should lead to a smaller diffusion coefficient and increased SR losses. As before, Figure~\ref{fig:tlsB_E} presents different timescales versus particle energy. For a lower $B$-field, the diffusion timescale is relatively shorter and particles will escape faster from a particular zone. Also, the SR timescale is longer and therefore it takes a longer time for particles to lose energy due to SR. 
An interesting regime change between IC and SR domination occurs around $\lesssim1$ TeV for these two $B$-field strengths: for $B=10\,\mu$G SR dominates over IC for $E_{\rm e}~\gtrsim~3~$TeV, while this change occurs around $E_{\rm e}\gtrsim10$~TeV for $B=3\,\mu$G. 

\begin{figure}
     \centering
     \includegraphics[height=5.5cm, width=8cm]{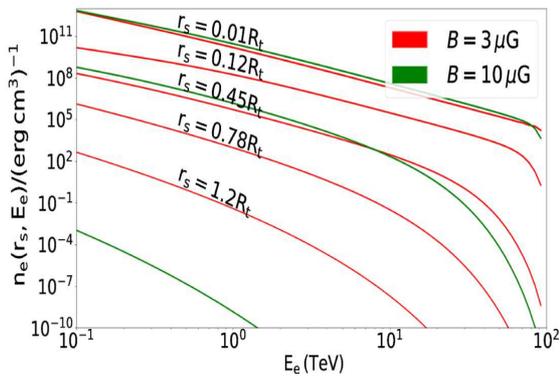}
    \caption{Steady-state particle spectrum as a function of energy for different $B$-fields. The labels (indicating different $r_{\rm s}$) are for the $B=3\mu$G case.}
     \label{fig:melBE}
   \end{figure}

\begin{figure}
    \centering
    \includegraphics[height=5.5cm, width=8cm]{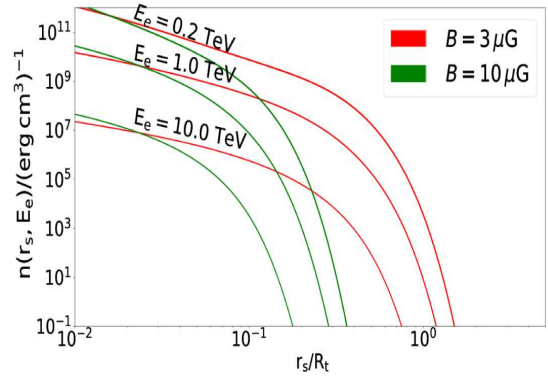}
    \caption{Steady-state particle spectrum versus radius for different $B$-fields.}
    \label{fig:melBr}
  \end{figure}

In Figure~\ref{fig:melBE}, at fixed radius, $n_{\rm e}$ is higher for lower $B$-field beyond $r_{\rm s}\gtrsim0.05R_{\rm t}$. This is because $\dot E_{\rm SR}$ is lower in this case and more particles survive. There is a cutoff at higher energies due to SR. As before, the cutoff becomes increasingly lower at larger radii. We note that the particle spectrum is very small at larger radii for $B = 10\,\mu\text{G}$ (i.e., the green line for the larger $B$-field is very low at $r_{\rm s}>0.5R_{\rm t}$ and thus not visible).
In Figure~\ref{fig:melBr}, at small radius the density is higher for higher $B$-field because of slower diffusion. However, at large $r_{\rm s}$, it is evident that a larger $B$-field leads to a substantially lower cutoff energy due to an increased $\dot{E}_{\rm SR}$ in this case.

\begin{figure}
   \centering
   \includegraphics[height=5.5cm, width=8cm]{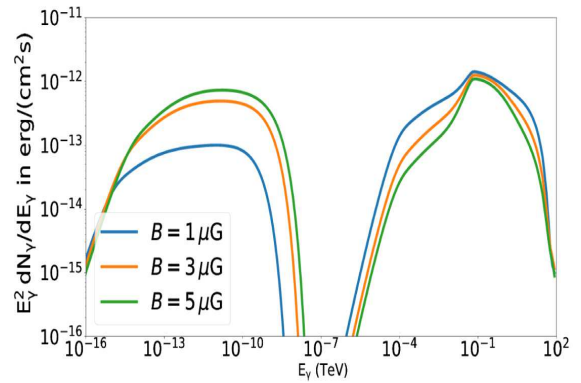}
  \caption{An SED plot indicating the effect of changing the $B$-field in the cluster.}
  \label{fig:B}
 \end{figure}

\begin{figure}
   \centering
   \includegraphics[height=5.5cm, width=8cm]{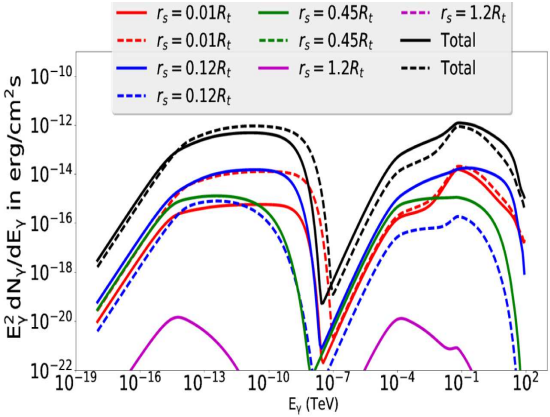}
  \caption{Spectra at different radii for different $B$-fields. The dashed lines represent $B=10~\mu$G and the solid lines represent $B=3~\mu$G.}
   \label{fig:dsp_B}
 \end{figure}

Figure~\ref{fig:B} shows the SED components of a cluster for different $B$-fields. The SR loss rate strongly depends on $B$. If we increase $B$ from $1\, \mu \text{G}$ to $5\, \mu \text{G}$ the SR losses increase rapidly because $\dot E_{\rm SR}\propto E^{2}B^{2}$, leading to a higher SR flux. In addition, higher-energy particles lose more energy leaving fewer of these particles to radiate IC at higher energies.
In Figure~\ref{fig:dsp_B}, we study the radial dependence of the cluster radiation. The dashed lines represent $B=10~\mu$G and solid lines represent $B=3~\mu$G. At the smallest radii ($r_{\rm s}\sim 0.01\rm R_{\rm t}$), both SR and IC emission increase with an increase in $B$-field because of slower diffusion in this case, leading to particles spending a longer time there and therefore radiating more efficiently. Different diffusion coefficients will lead to different source sizes. If particles move faster the source will be larger. Thus a higher $B$-field leads to a more compact source. This explains the line swap in the graph (around $r_{\rm s}\sim 0.1\rm R_{\rm t}$), i.e., the non-monotonic behaviour of lines as $r_{\rm s}$ is increased (cf.\ Figure $6$ from \citealt{Kopp2013}). Both IC and SR fluxes decrease with an increase in $B$-field at large radii (making the green and magenta dashed lines invisible on Figure~\ref{fig:dsp_B}).

\subsection{Changing the electron source term: Spectral index $\Gamma$}
\begin{figure}
   \centering
   \includegraphics[height=5.5cm, width=8cm]{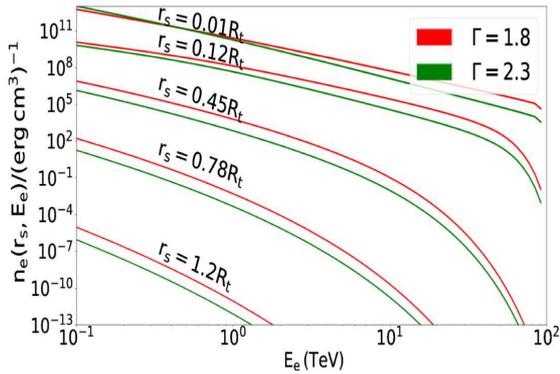} 
  \caption{Steady-state particle spectrum as a function of energy at different radii for two different $\Gamma$.}
   \label{fig:spectral_energy}
 \end{figure}

\begin{figure}
   \centering
   \includegraphics[height=5.5cm, width=8cm]{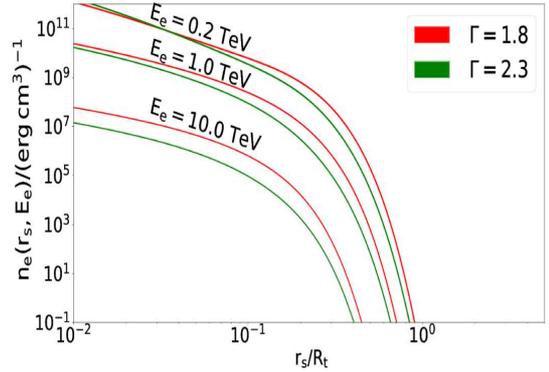}
   \caption{Steady-state particle spectrum as a function of radius at different energies for two different $\Gamma$.}
   \label{fig:spectral_radius}
 \end{figure}

Changing $\Gamma$ does not affect the timescales versus $E_{\rm e}\,{\rm or}\,r_{\rm s}$. However, it changes the source term (see Eq.~[\ref{eq:10}], where we change $\Gamma$ and keep $L_{\rm e}$ constant). 
Figure~\ref{fig:spectral_energy} shows the steady-state particle spectrum as a function of energy for different spectral indices $\Gamma$ as denoted by the colours in the legend. At a fixed radius, the steady-state particle spectrum $n_{\rm e}$ is higher for a harder injection spectrum at large radii and energies. However, close to the core the low-energy tail of the spectrum is slightly higher for a softer spectral index because $L_{\rm e}$ is kept constant (i.e., we assume the same total power in particles). At low energies and smaller radii the effect of changing the $\Gamma$ is smaller. This can also be seen in Figure~\ref{fig:spectral_radius}.
 
\begin{figure}
   \centering
   \includegraphics[height=5.5cm, width=8cm]{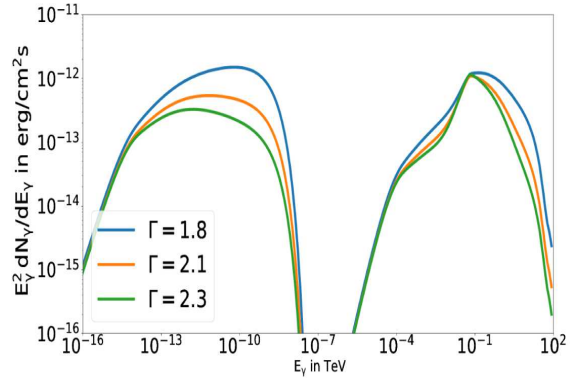}
 \caption{An SED plot indicating the effect of changing the spectral index on the SR and IC components.}
 \label{G}
 \end{figure}

Figure~\ref{G} shows the SED plot in the case of changing $\Gamma$. Both SR and IC decline for a soft $\Gamma$ and increase for a harder injection spectrum (since more high-energy particles are injected into the GC in this case).

\subsection{Changing the electron source term: Normalisation $Q_{0}$}
Changing the source term does not affect the timescales versus $E_{\rm e}\,{\rm or}\,r_{\rm s}$, but simply increases or reduces the number of particles injected into the cluster, and thus the level of radiation received from the cluster. Thus the SED scales linearly with $Q_0$.
 
\subsection{Changing the source distance $d$}
Changing the distance to the cluster changes the physical size of the cluster since we keep the angular size constant (e.g., the core radius $r_{\rm c} = \theta_{\rm c} d$, with $\theta_{\rm c}$ the `angular radius' of the core). For a smaller distance the IC timescale is lower (i.e., the volume is smaller and thus the energy density becomes larger, leading to an increased IC loss rate). Furthermore, since $t_{\rm diff}\propto r_{\rm s}^{2}\propto d^{2}$, when the distance decreases particles escape faster from a particular zone, leading to lower diffusion timescales. However, the SR timescale is not influenced by a change in distance, because we assumed a constant $B$-field everywhere in the cluster. 

\begin{figure}
   \centering
   \includegraphics[height=5.5cm, width=8cm]{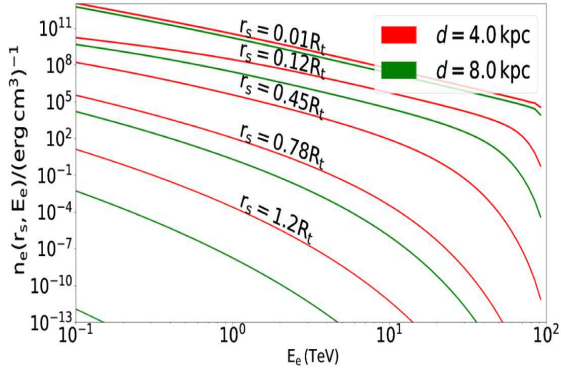}
   \caption{Steady-state particle spectrum as a function of energy for different distances to the cluster (see legend).}
   \label{fig:meldE}
 \end{figure}

\begin{figure}
   \centering
   \includegraphics[height=5.5cm, width=8cm]{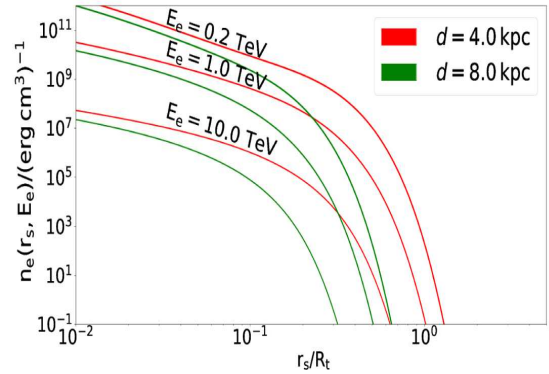}
   \caption{Steady-state particle spectrum as a function of radius for different distances (see legend).}
   \label{fig:meldr}
 \end{figure}

\begin{figure}
 \centering
   \includegraphics[height=5.5cm, width=8cm]{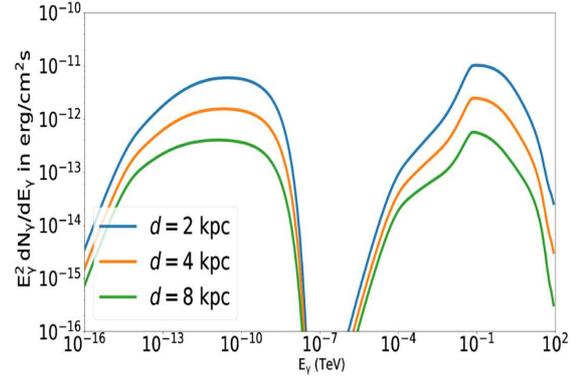}
 \caption{An SED plot showing the effect of changing the distance to the cluster on SR and IC components.}
 \label{D}
 \end{figure}

In Figure~\ref{fig:meldE} one can see that the steady-state particle spectrum (per volume) is higher for a smaller distance due to a decrease in volume $\propto d^{3}$.
The effect of changing the distance to the cluster is lower at small radii (see also Figure~\ref{fig:meldr}). We used the same number of zones for these two cases of $d$, but the actual $R_{\rm t}\propto d$ changes. Thus the relative radius $r_{\rm s}/R_{\rm t}$ changes. Physically, the green lines are associated with a larger cluster than the red lines. Thus particles take longer to escape (having to traverse a longer actual path) and therefore lose more energy via SR losses while diffusing through the cluster. This explains the lower spectral cutoff as a function of $r_{\rm s}/R_{\rm t}$ for a larger $d$.
Figure~\ref{D} shows that as the distance to the cluster increases from $2~\rm{kpc}$ to $8~\rm{kpc}$, both the SR and IC fluxes decrease. This is due to a lower IC loss rate, and the fact that particles diffusing through clusters that are more distant (larger physical source) tend to lose more energy via SR. Fluxes furthermore scale as $1/d^{2}$. All of these effects suppress the flux when $d$ is increased. 

\subsection{Changing the spatial diffusion coefficient $\kappa(E)$}
\begin{figure}
   \centering
   \includegraphics[height=5.5cm, width=8cm]{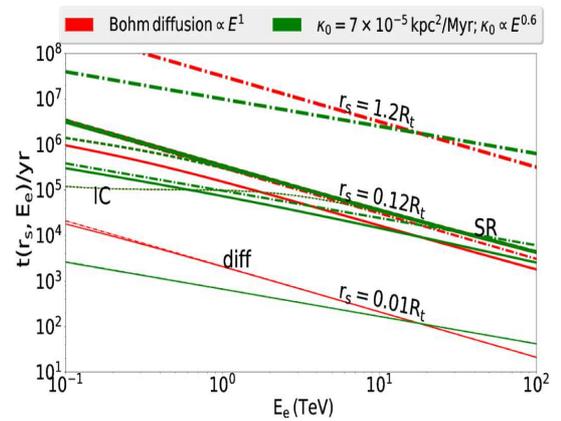}
   \caption{Timescales as a function of energy for different diffusion coefficients.}
   \label{fig:tlsKE}
   \end{figure}

In Figure~\ref{fig:tlsKE} we plot the timescale graph for diffusion (dash-dotted lines), radiation losses (dashed lines), and effective timescale (solid lines). We used two diffusion coefficients, Bohm diffusion $\kappa \sim 6\times10^{24}\,\rm cm^{2}/s\sim2\times10^{-5}{\rm kpc^{2}/Myr}$ at 1 TeV that goes like $\kappa\propto E_{\rm e}^{1}$, and $\kappa_{0}\sim 2\times10^{25}\,\rm cm^{2}/s\sim7\times10^{-5}{\rm kpc^{2}/Myr}$ at 1 TeV that goes like $\kappa\propto E_{\rm e}^{0.6}$. The colours represent these two diffusion coefficients as shown in the legend. High-energy particles diffuse faster than low-energy particles (given the negative slopes of the diffusion timescales). The Bohm diffusion timescale therefore has a slope of $-1$ while in the second case, $t_{\rm diff}$ has a slope of $-0.6$. The Bohm diffusion timescale is larger at $E_{\rm e}\lesssim20\,\rm TeV$ and smaller at $E_{\rm e}\gtrsim20\,\rm TeV$. The radiation timescales, however, do not change.

\begin{figure}
   \centering
   \includegraphics[height=5.5cm, width=8cm]{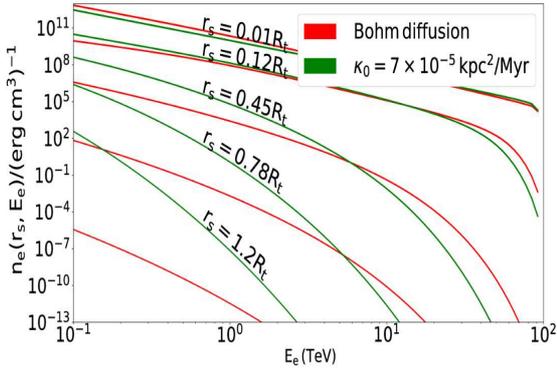}
   \caption{Steady-state particle spectrum as a function of energy for different diffusion coefficients.}
   \label{fig:melKE}
 \end{figure}

\begin{figure}
   \centering
   \includegraphics[height=5.5cm, width=8cm]{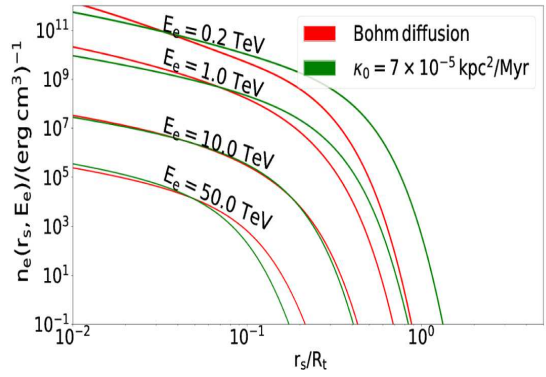}
   \caption{Steady-state particle spectrum as a function of radius for different diffusion coefficients.}
   \label{fig:melKr}
 \end{figure}

In Figure~\ref{fig:melKE}, at a fixed radius, $n_{\rm e}$ is higher for the $\kappa\propto E_{\rm e}^{0.6}$ value of the diffusion coefficient at low energies at $r_{\rm s}\gtrsim0.45R_{\rm t}$ (at the lowest radii diffusion dominates and $n_{\rm e}$ is slightly higher for the Bohm case). More particles reach larger radii (outer zones) due to faster diffusion. At VHEs, the Bohm diffusion is now faster, and thus there are slightly fewer particles in the centre of the cluster. The $n_{\rm e}$ is therefore higher at larger radii and energies and vice versa. One can observe the same effect in Figure~\ref{fig:melKr}: $n_{\rm e}$ is slightly higher for the Bohm case at the lowest radii and low energies, and drops below the value of the second case at larger radii, but exceeds that value again for the highest energies.
As particles diffuse out from the centre of the cluster, they lose energy, e.g., 20 TeV particles become 10 TeV particles, and thus the cross-over point where the two values of $n_{\rm e}$ coincides moves toward increasingly lower energies (i.e., it is a cooling effect).

\begin{figure}
 \centering
   \includegraphics[height=5.5cm, width=8cm]{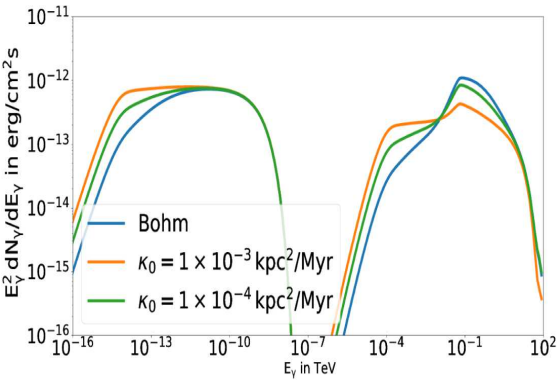}
 \caption{SED plot for a cluster with a change in spatial diffusion coefficient.}
 \label{kappa}
 \end{figure}

In Figure~\ref{kappa} we see that the high-energy IC component is higher for Bohm diffusion. This is because this component originates mostly at the GC centre, and there are more particles in the Bohm case. Since particles lose more energy due to IC for slower diffusion, there is less energy available for SR, and thus the SR flux scales with the spatial diffusion coefficient at low energies (since $B$ is not a function of $r$). 

\begin{figure}
   \centering
   \includegraphics[height=5.5cm, width=8cm]{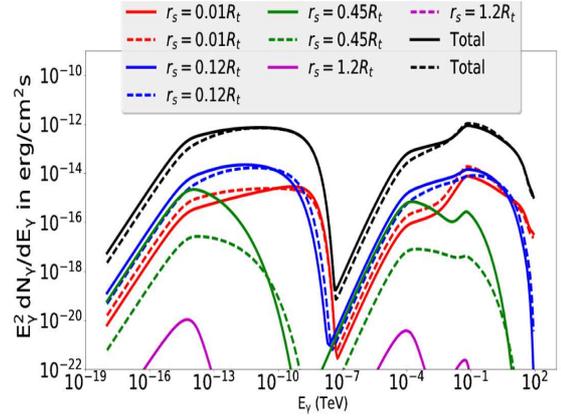}
  \caption{Spectra at different radii upon changing the diffusion coefficients. The dashed lines represent Bohm diffusion and the solid lines represent $\kappa_{0}=7\times10^{-5}{\rm kpc^{2}/Myr}$.} 
   \label{fig:dsp_kappa}
 \end{figure}

In Figure~\ref{fig:dsp_kappa} we see that the optical IC is dominated by emission from particles at the centre of the cluster, and thus is higher for Bohm diffusion. Bohm diffusion (dashed lines) is relatively slower, therefore there are fewer particles at larger radii, leading to the line swap as radius is increased. At VHEs, there are few particles and hence the cutoff due to the Klein-Nishina effect is evident.

\subsection{Summary}
Table \ref{parameter_summary} summarises the effect of increasing a particular model parameter with respect to its reference value (Table~\ref{parameter_table}) on the particle spectrum. Also, Table~\ref{SED_summary} summarises the effect of such changes on the predicted SED. 
 
\begin{table*}
  	\caption{The effect of increasing model parameters with respect to their reference values on the steady-state particle spectrum.}
  	\label{parameter_summary}
  	\centering
  	\begin{tabular}{|c | c | c | c | c |}
  \hline
 \textbf{Model parameters} & \multicolumn{2}{c|}{\textbf{Close to the core}} & \multicolumn{2}{c|}{\textbf {$ r\approx R_{\rm t}$}}\\
 \cline{2-5}
 & \textbf{Low energies} & \textbf{High energies} & \textbf{Low energies} & \textbf{High energies} \\ 
 \hline
 Number of stars ($N_{\rm tot}$)		 & - & = & = & = \\
 Magnetic field ($B$)       		 & = & $\approx$ & - & - \\
 Spectral index ($\Gamma$)      		 & $\approx$ & - & - & - \\
 Electron source term ($Q_{0}$) 		 & + & + & + & + \\
 Distance ($d$)                			 & - & - & - & -\\
 Diffusion coefficient ($\kappa_{0}$) 		 & $\approx$ & = & + & + \\
 	\hline
 \end{tabular}
\end{table*}

 \begin{table*}
 	\caption{The effect of increasing model parameters with respect to their reference values on the predicted SED.}
 	\label{SED_summary}
 	\centering
 	\begin{tabular}{|c | c | c | c | c | c | c |}
 		\hline
 	 \textbf{Model parameters} & \multicolumn{3}{c|}{\textbf{Close to the core}} & \multicolumn{3}{c|}{\textbf{$ r\approx R_{\rm t}$}} \\ 
 	 \cline{2-7}
 	 & \textbf{Radio} & \textbf{X-rays} & \textbf{$\gamma$-rays} & \textbf{Radio} & \textbf{X-rays} & \textbf{$\gamma$-rays} \\
 	 \hline
 	 Number of stars ($N_{\rm tot}$)		 & - & = & + & $\approx$ & $\approx$ & $\approx$ \\
 	 Magnetic field ($B$)           		 & + & + & $\approx$ & - & - & - \\
 	 Spectral index ($\Gamma$)      		 & $\approx$ & - & - & - & - & - \\
 	 Electron source term ($Q_{0}$) 		 & + & + & + & + & + & + \\
 	 Distance ($d$)                		 & - & - & - & - & - & - \\
 	 Diffusion coefficient ($\kappa_{0}$) & + & + & $\approx$ & - & - & $\approx$  \\	
 \hline
\end{tabular}
\end{table*}

\section{Spectral Modelling of Selected Globular Clusters}
\label{Spectral}
\citet{Abramowski2013} performed an analysis of H.E.S.S.\ data to search for VHE emission from 15 GCs. They could not detect any individual cluster. They also performed a stacking analysis, but even in this case there was no significant cumulative signal. This means that Terzan~5 is the only Galactic GC plausibly detected at VHEs \citep{Abramowski2011}. We list some structural parameters of the analysed GCs in Table~\ref{table:1}. In what follows we use the VHE upper limits of the 15 GCs and a measured spectrum of Terzan~5. In addition, diffuse X-ray emission has been detected from two of these GCs: Terzan~5 \citep{Eger2010}, and 47~Tucanae \citep{Wu2014}; an upper limit has been obtained for NGC 6388 \citep{Eger2012}. The measured X-ray spectra are, however, much harder than what is predicted by our model. We therefore postulate that this points to a new spectral component that we have not modelled yet. Therefore we treat these X-ray data as upper limits for our SR component (Section~\ref{sec:constraining}). Finally, in Section~\ref{SEDs} we apply our model to all 16 GCs, for fixed parameters, and obtain a ranking according to the predicted VHE flux for both H.E.S.S.\ and CTA. We also list the five most promising GCs according to their predicted VHE flux. 

\subsection{Structural and Other Characteristics of Selected GCs}
\label{sec:Characteristics_GCs}
Fifteen Galactic GCs were selected by \citet{Abramowski2013} for analysis by applying \textit{a priori} cuts on the target and observational run lists of H.E.S.S.: the GCs lie off the Galactic Plane by more than $1.0^\circ$, the pointing position was within $2.0^\circ$ of the GC position, and at least 20 high quality runs were available for each source. We list some structural parameters in Table~\ref{table:1}: the core ($R_{\rm c}$), tidal\footnote{\url{http://gclusters.altervista.org/}} ($R_{\rm t}$), and half-mass ($R_{\rm h}$) radius (\citealt{Harris1996,Harris2010}) and the number of stars ($N_{\rm tot}$) estimated from $N_{\rm tot} = 10^{0.4(4.79-M_{V})}$ hosted by each cluster, where $M_{V}$ is the integrated absolute magnitude \citep{Lang1992}, the estimated number of MSPs in each GC \citep{Abdo2010,Venter2015}, and their distances from the Sun \citep{Harris2010}. Furthermore we use the updated value of $R_{\rm c} = 5.94\arcsec = 0.1\arcmin$ for Terzan~5 as given by \citet{Prager2016}. We also list the assumed values for $Q_{0}$.

\begin{table*}
\caption{Structural characteristics and other parameters of selected GCs.}
\label{table:1}
\centering 
\begin{tabular}{|c | c | p{1cm} | c | c | c | c | c |}          
\hline                      
GC name & $R_{\rm c}$(pc) & $R_{\rm hm}$~(pc) & 
$R_{\rm t}$ (pc) & Estimated $\rm N_{\rm tot}/10^{5}$ & $N_{\rm MSP}$ & $d~(\rm kpc)$ & \vtop{\hbox{\strut  $Q_{0}$ 
   \hbox{\strut($\times10^{33}\rm erg^{-1}s^{-1})$}}}\\     
\hline                                   
    NGC 104    & 0.47   & 4.1   & 56.1 & 4.57 & $>$ 33 & 4.5 & $9.55$\\
    NGC 6388   & 0.16   & 0.68  & 8.13 & 5.81 & 100 & 9.9 &  $52.1$\\
    NGC 7078   & 0.18   & 1.31  & 28.1 & 4.13 &  25 & 10.4 & $7.24$\\
    Terzan 6   & 0.07   & 0.58  & 22.8 & 0.29 &  25 & 6.8 & $7.24$\\
    Terzan 10  & 1.18   & 2.03  & 6.62 & 0.30 &  25 & 5.8 & $7.24$\\
    NGC 6715   & 0.12   & 1.07  & 9.78 & 4.79 &  25 & 26.5 & $7.24$\\  
    NGC 362    & 0.24   & 1.07  & 21.1 & 1.58 &  25 & 8.6 & $7.24$\\
    Pal 6      & 0.86   & 1.57  & 10.9 & 0.31 &  25 & 5.8 & $7.24$\\
    NGC 6256   & 0.03   & 1.13  & 9.94 & 0.21 &  25 & 10.3 & $7.24$\\
    Djorg 2    & 0.43   & 1.31  & 13.8 & 0.51 &  25 & 6.3 & $7.24$\\
    NGC 6749   & 0.81   & 1.44  & 6.82 & 1.78 &  25 & 7.9 & $7.24$\\
    NGC 6144   & 1.23   & 2.13  & 43.5 & 0.48 &  25 & 8.9 & $7.24$\\
    NGC 288    & 1.77   & 2.92  & 16.9 & 0.32 &  25 & 8.9 & $7.24$\\
    HP 1       & 0.04   & 4.06  & 10.8 & 0.48 &  25 & 8.2 & $7.24$\\
    Terzan 9   & 0.04   & 1.02  & 10.8 & 0.02 &  25 & 7.1 & $7.24$\\
    Terzan 5   & 0.17   & 0.94  & 17.4 & 0.77  &  $>$ 34 & 5.9 & $20.0$\\
\hline                                             
\end{tabular}
\end{table*}

\subsection{Constraining Parameters via X-ray and $\gamma$-ray Data}
\label{sec:constraining}
\begin{figure}
   \centering
   \includegraphics[height=5.5cm, width=8cm]{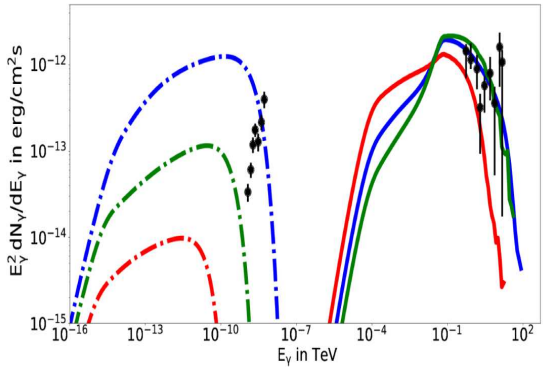}
  \caption{The predicted SED for Terzan~5 indicating the SR (integrated between $55^\prime<r_{\rm s}<174^\prime$, the dash-dotted lines) for the inner part of the source and IC (integrated over all $r_{\rm s}$, solid lines) components for different combinations of parameters (Table~\ref{table:2}), as well as \emph{Chandra} \citep{Eger2010} and H.E.S.S.\ \citep{Abramowski2011} data.}
  \label{terzan5}  
 \end{figure}

\begin{figure}
   \centering
    \includegraphics[height=5.5cm, width=8cm]{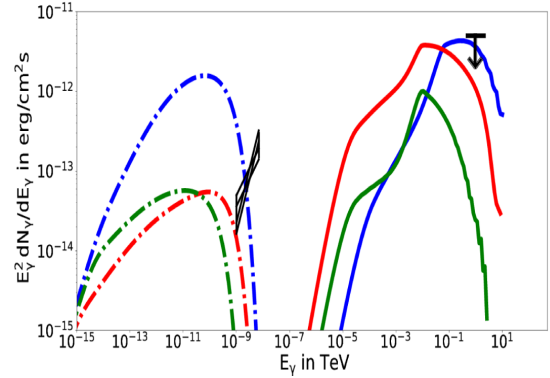}
     \caption{The predicted SED for 47 Tucanae indicating the SR (integrated between $0.72^\prime<r_{\rm s}<3.835^\prime$, the dash-dotted lines), and IC (integrated over all $r_{\rm s}$, solid lines) components for different combinations of parameters (Table~\ref{table:3}), as well as \emph{Chandra} data \citep{Wu2014} and a H.E.S.S.\ upper limit \citep{Abramowski2013}.}
          \label{tucanae}
 \end{figure}

\begin{figure}
  \centering
  \includegraphics[height=5.5cm, width=8cm]{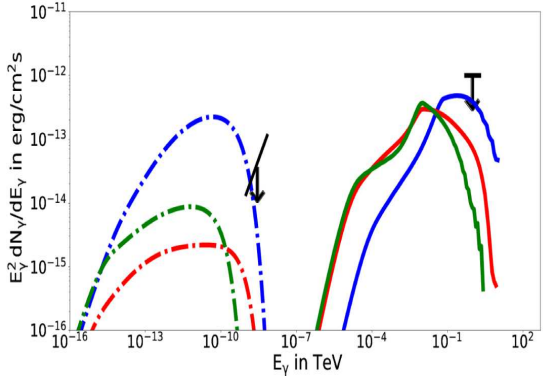}
    \caption{The predicted SED for NGC~6388 indicating the SR (integrated between $25''<~r_{\rm s}<~139''$, the dash-dotted lines) and IC (integrated over all $r_{\rm s}$, solid lines) components for different combinations of parameters (Table~\ref{table:4}). The power law and arrow represent the X-ray upper limit \citep{Eger2012} whilst the line-arrow represents the H.E.S.S.\ upper limit \citep{Abramowski2013}.}
         \label{NGC6388}
 \end{figure}

We used diffuse X-ray and VHE\footnote{We do not use \emph{Fermi} LAT data since we do not model the cumulative pulsed curvature emission as was done by, e.g., \citet{Venter2009a}.} $\gamma$-ray observations to constrain cluster parameters for three sources (i.e., Terzan~5, 47~Tucanae, and NGC~6388). We first present the results for Terzan~5. The $\gamma$-ray \citep{Abramowski2011} and X-ray \citep{Eger2010} data are plotted in Figure~\ref{terzan5}. Our model cannot reproduce the flat slope of the X-ray data. Hence, we postulate a new radiation component (see Venter et al., in preparation, who attribute this to cumulative pulsed SR from the individual MSP magnetospheres) to explain these data. We therefore treat the X-ray data as upper limits and our predicted SR component must be below these. We are not fitting the data using rigorous statistical techniques, but we are simply trying to find sample parameters so that our predicted SR component is not in conflict with the X-ray ``upper limits'', while still fitting the $\gamma$-ray data. Figure~\ref{terzan5} shows the predicted differential SED components for Terzan~5 indicating the predicted SR (integrated between $55^\prime<r_{\rm s}<174^\prime$\footnote{In Figures~\ref{terzan5}, \ref{tucanae}, and \ref{NGC6388} the dash-dotted lines represent the inner part of the source, and the solid lines indicate the whole source visible in $\gamma$-rays. The field of view (FoV) of H.E.S.S.\ is so large that one can see the whole source, while only a small part of the source is seen in X-rays, since the FoV of \emph{Chandra} is relatively small.} corresponding to the X-ray data, \citealt{Eger2010}) and IC (integrated over all $r_{\rm s}$) components using a combination of parameters so as not to violate the \textit{Chandra} and H.E.S.S.\ data. We show the blue component as an example of a parameter combination which violates the data. On the other hand, both the green and the red components satisfy the data. See Table~\ref{table:2} for the parameters corresponding to these lines. Thus one can actually allow\footnote{We could not constrain the radio and optical part of the spectrum because of lack of data. There might be some radio data available \citep{Clapson2011}, but it is not certain that these data are associated with the diffuse SR component that we predict. In this work we mainly focus on $\gamma$-ray and X-ray data.
The predicted optical diffuse flux level is extremely low and that makes it very difficult to find upper limits or data that would at all be constraining, given the number of optical sources ($N_{\rm tot}$) that dominate the radiation in this band.}
$E_{\rm e, max}$ to be $\sim30~\rm TeV$ to $100~\rm TeV$ and the $B$-field can be $\sim1~\mu G$ to $4~\mu G$, depending on other parameters. Other combinations of parameters also exist that can give similar fits. One should be able to break this degeneracy by adding more data in future.

 \begin{table*}
 	\caption{Parameter combinations for each of the line colours appearing in Figure~\ref{terzan5} for Terzan~5.}               
 	\label{table:2}     
 	\centering     

 	\begin{tabular}{|c | c | p{1cm} | c | c | c | c | c |}          
 		\hline      

 		Line colours & $\kappa~({\rm kpc^{2}/Myr})$ & $B~(\mu\rm{G})$ & $\Gamma$ & $Q_{0}~({\rm erg^{-1}s^{-1}})$ & $d$~(kpc) & $N_{\rm tot}$ & $E_{\rm e, max}$ (TeV) \\
 		\hline
 		
 		Blue 	&	Bohm diffusion	& 5	& 1.8 & $1.16\times10^{34}$ & 5.9 & $7.7\times10^4$ & 100 \\
 		Red  	&	Bohm diffusion  & 1 & 1.8 & $6.33\times10^{33}$ & 5.9 & $7.7\times10^4$ & 20  \\
 		Green	&	$0.7\times10^{-4}$ & 2 & 2.0 & $9.84\times10^{33}$ & 5.9 & $7.7\times10^4$ & 50\\
\hline                                             
\end{tabular}
\end{table*}	

Similarly, we present results for 47~Tucanae using diffuse X-ray data \citep{Wu2014} and H.E.S.S.\ upper limits \citep{Abramowski2013} to constrain model parameters. Figure~\ref{tucanae} shows the predicted SED for 47~Tucanae indicating the predicted SR (integrated between $0.72^\prime<r_{\rm s}<3.835^\prime$) and IC (integrated over all $r_{\rm s}$) components using several parameter combinations. The blue line violates the X-ray data (treated as ``upper limits''). The parameters used for each line are summarised in Table~\ref{table:3}. We find for example that $E_{\rm e, max}\lesssim100\,{\rm TeV}\,{\rm and}\,B\,\lesssim\,5\,\mu G$ satisfy the measurements. Thus, the degeneracy of parameters is again evident. 

\begin{table*}
	\caption{Parameter combinations for each of the line colours appearing in Figure~\ref{tucanae} for 47~Tucanae.}               
	\label{table:3}     
	\centering                                      
	\begin{tabular}{|c | c | p{1cm} | c | c | c | c | c |}          
		\hline      
		Line colours & $\kappa~({\rm kpc^{2}/Myr})$ & $B~(\mu\rm{G})$ & $\Gamma$ & $Q_{0}~({\rm erg^{-1}s^{-1}})$ & $d$~(kpc) & $N_{\rm tot}$ & $E_{\rm e, max}$ (TeV) \\
		\hline
		
		Blue 	&	Bohm diffusion	& 5	& 1.8 & $1.16\times10^{34}$ & 4.5 & $4.6\times10^5$ & 10 \\
		Red  	&	$1.1\times10^{-4}$ & 1 & 2.0 & $9.55\times10^{33}$ & 4.5 & $4.6\times10^5$ & 100 \\
		Green	&	$1.1\times10^{-4}$ & 4 & 2.3 & $3.18\times10^{33}$ & 4.5 & $4.6\times10^5$ & 30\\
		\hline                                          
\end{tabular}
\end{table*}	

Figure~\ref{NGC6388} shows the predicted SED for NGC 6388. The parameter values used are summarised in Table~\ref{table:4}. Again, we note that there are different combinations of parameters that satisfy the data constraints, e.g., we require a small $B$ and $E_{\rm e, max}$, or a small $Q_{0}$ and large $\Gamma$ to satisfy the \emph{Chandra} and H.E.S.S.\ upper limits.

 \begin{table*}
 	\caption{Parameter combinations for each of the line colours appearing in Figure~\ref{NGC6388} for NGC~6388.}               
 	\label{table:4}     
 	\centering                                      
 	\begin{tabular}{|c | c | p{1cm} | c | c | c | c | c |}          
 		\hline      
 		Line colours & $\kappa~({\rm kpc^{2}/Myr})$ & $B~(\mu\rm{G})$ & $\Gamma$ & $Q_{0}~({\rm erg^{-1}s^{-1}})$ & $d$~(kpc) & $N_{\rm tot}$ & $E_{\rm e, max}$ (TeV) \\
 		\hline
 		
 		Blue 	&	Bohm diffusion	& 5	& 1.8 & $6.33\times10^{33}$ & 9.9 & $5.8\times10^5$ & 10 \\
 		Red  	&	Bohm diffusion & 1 & 2.0 & $3.47\times10^{33}$ & 9.9 & $5.8\times10^5$ & 100 \\
 		Green	&	$1.1\times10^{-4}$ & 2 & 2.3 & $5.21\times10^{33}$ & 9.9 & $5.8\times10^5$ & 30\\
 		\hline
 	\end{tabular}
 \end{table*}
 
 \subsection{Ranking the GCs according to predicted VHE flux}
 \label{SEDs}
\begin{figure}
  \centering
  \includegraphics[height=5.5cm, width=8cm]{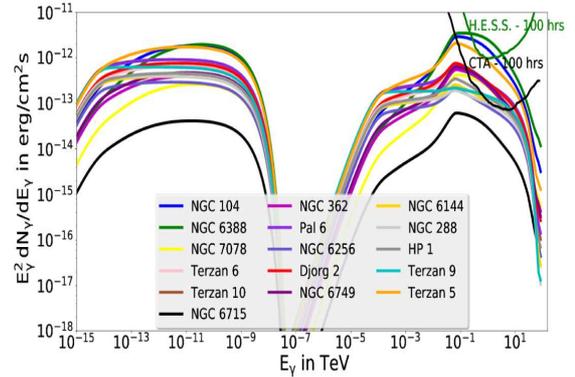}
  \caption{Predicted differential spectra ${E^{2}_{\gamma}} \, {dN_{\gamma}/dE_{\gamma}}$ in ${\rm ergcm^{-2}s^{-1}}$ for 15 non-detected GCs and for Terzan~5. The two components represent the SR and IC spectra. The H.E.S.S.\ and CTA sensitivities (for 100 hours) are also shown.}
  \label{GCs}
  \end{figure}

We apply the model described in Section~\ref{model} to 15 non-detected GCs and to Terzan~5 using the fixed parameters and value of $Q_{0}$ given in Table~\ref{table:1}. We assume Bohm diffusion, $\Gamma = 2.0$, $B = 5\,\mu \text{G}$, and $E_{\rm e, max}~=~100$~TeV as a reference to produce SR and IC spectra for each individual cluster. According to our flux predictions, H.E.S.S.\ may detect two more GCs, i.e., 47~Tucanae (blue) and NGC 6388 (green) in addition to Terzan~5 (orange) if the clusters are observed for 100 hours (see Figure~\ref{GCs}). The clusters 47~Tucanae and NGC~6388 have not been detected by H.E.S.S.\ yet but they have only been observed for about 20 hours each. We note, however, that this flux prediction and therefore the ranking is very sensitive to the choice of parameters (implying significant error bars on the predicted fluxes). The CTA will be 10 times more sensitive than H.E.S.S.\ and therefore may detect many more GCs. We find that more than half of the known Galactic population may be detectable for CTA, depending on observation time and model parameters. However if there is a non-detection by CTA, this will imply strong parameter constraints or even model constraints (i.e., the model might be not viable any more). The top five most promising GCs for CTA are NGC~6388, 47~Tucanae, Terzan~5, Djorg~2, and Terzan~10 as seen in Figure~\ref{GCs}.

\section{Discussion and Conclusion}
\label{Conclusion}
This paper focused on constraining model parameters for Galactic GCs using $\gamma$-ray and X-ray data, with the main aim being to study the detectability of GCs for H.E.S.S.\ and CTA. We used a leptonic emission code to make flux predictions and performed a parameter study, varying six model parameters. 		
For Terzan~5 we found that $E_{\rm e, max}$ could be varied between 30 TeV and 100 TeV and the $B$-field between $1~\mu G$ and $4~\mu G$ while still fitting the SED by fixing the other parameters. Similarly we found that for 47~Tucanae, $E_{\rm e, max}\leq~100$ TeV and $B~<~5~\mu G$, and we require a small $B$ and $E_{\rm e, max}$ or a small $Q_{0}$ and large $\Gamma$ for NGC~6388 in order to satisfy the upper limits. We therefore found that the parameters of the individual GCs were uncertain and quite unconstrained by the available data, and we noted that there were different combinations of parameters that satisfied the observational constraints (i.e., they were degenerate). We also found that the predicted IC component for the majority of the 16 GCs we studied were below the H.E.S.S. sensitivity limit. However, H.E.S.S.\ may detect two more GCs, i.e., 47 Tucanae and NGC 6388, if the clusters are observed for 100 hours. On the other hand, CTA may detect many more GCs (possibly more than half of the known Galactic population, depending on observation time and model parameters). The five most promising GCs are NGC~6388, 47~Tucanae, Terzan~5, Djorg~2, and Terzan~10. Future multi-wavelength studies should allow us to constrain some parameters better as well as discriminate between competing radiation models. 

 \bibliographystyle{mnras}
 \bibliography{references}

\label{lastpage}
\end{document}